Editors' Suggestion

# Physics computational literacy: Programming, modeling, and collaboration at the journeyman level

Karl Henrik Fredly,[1,*] Tor Ole B. Odden,[1] and Benjamin M. Zwickl[1,2]

[1]*Center for Computing in Science Education, University of Oslo, 0316 Oslo, Norway*

[2]*School of Physics and Astronomy, Rochester Institute of Technology, 84 Lomb Memorial Drive, Rochester, New York 14623, USA*

Computation has become an integral part of physics research. However, little is known about how students learn to productively use computation as a tool beyond the introductory level, especially as they transition into physics research. In this study, we apply the theory of physics computational literacy and the novice-expert framework to describe the development of expertise in computational physics, as students transition from novice to journeyman computational physicists. We base this description on a thematic analysis of interviews with 13 computational physics master's students with extensive experience using computation. We first describe the most important elements driving the development of computational physics expertise, identifying two distinct transitions of competence during their studies, driven by experience with large computational projects and professional research. We then present an overview of the various skills students attain on this path toward the journeyman level of computational physics expertise. Based on these results, we argue for the need to assist students in collaborative coding and in the learning of new tools, as well as for the importance of large, scaffolded, computational projects in helping students develop the advanced skills needed for computational research.

## I. INTRODUCTION

Computation is critical to the field of physics, both for research and industry applications. Programming is now widely used by physics Ph.D. students [1], and one of the top categories for jobs of physics bachelor's students is computing software [2]; more than half of physics graduates working in science, technology, engineering, and mathematics now use programming [3]. Even in several specialized physics related fields, programming is essential [4,5].

Computation has also become an important part of physics education [6]. Many physics students now learn both programming and numerical methods in their studies, skills that enhance physics learning [7] and prepare students for careers in industry [8].

In response to these trends, some universities now offer physics degree programs, minors, or specialized courses in computational physics [9–11]. Here, students not only learn advanced numerical methods but also develop advanced programming skills, such as high-performance computing, object-oriented programming, the use of code-collaboration tools, and version control. These skills are essential for professional computational work but are not commonly taught or required in introductory courses.

Although computational physics programs and courses are becoming more common, most theoretical frameworks for describing computational learning outcomes in physics are based on introductory courses. There has been relatively little work on applying these frameworks to study intermediate or advanced computational physics education. This represents a major gap in the literature, especially given the fact that most students learn computation in order to use it in research or industry careers.

In this study, we address this gap by examining how computational physics students at the master's level work on computational problems and how they develop their computational skills through their studies. To do so, we leverage the theory of physics computational literacy [12], which we elaborate based on interviews with computational physics students at a journeyman stage of computational physics expertise, wherein they have recently completed master's-level coursework and a computational physics research project.

In what follows, we provide a brief overview of the literature on computational physics education to clarify and situate this gap and then argue for the usefulness of computational literacy as a theoretical framework to address it. Next, we outline the context for our study:

*Contact author: karlhf@uio.no

the computational physics master's degree program at the University of Oslo. Based on a thematic analysis of semistructured interviews with a selection of these students, in the Results section, we describe how these students transitioned from novice to journeyman computational physics students and the elements of physics computational literacy that define this journeyman level of expertise. Based on these results, we conclude with implications for how one might teach these elements, such as the importance of computational projects in helping students go beyond novice levels of expertise.

## II. LITERATURE REVIEW

Computation has a long and rich history in physics education, going back almost to the founding of the standardized physics curriculum in the United States during the 1960s [13–16]. After the advent of the field of physics education research (PER) in the 1980s, these efforts became a regular area of study within PER. For instance, one of the early studies on the use of computation in physics education came from Redish and Wilson [17] in 1993, who studied the integration of programming into an introductory physics course using interactive and editable sample programs in Pascal. As programming has become more widespread, and as the programming tools themselves have developed, so has the literature.

Odden and Caballero [7] provide an overview of the literature on computing in physics education. According to their review, there are three main strands of research in this area. The first of these is the *modeling* strand. This strand builds on the long tradition in physics of modeling physical phenomena and uses computational tools to create computational physics models which are then used to learn key physics concepts and ideas. This strand is the most prevalent in the literature, and, relevant to the present study, it highlights the importance of *scaffolding* for teaching computational modeling techniques, such as minimally working programs [18]. This body of literature also takes into consideration how computation can be integrated with group work and standard physics assessment [19].

Second is the strand of research on *computational thinking (CT)*, a perspective that can be traced back to foundational work by Papert [20] and Wing [21,22]. While Wing's definition of CT focuses on "thinking like a computer scientist," other research takes a broader view of CT, where code and computers are unnecessary for CT [23]. Weintrop *et al.* [24], adapted CT to science education by creating a taxonomy of computational thinking practices in high school mathematics and science classrooms. It includes data practices such as collecting and visualizing data, modeling practices such as constructing computational models, and practices that are more focused on programming, such as troubleshooting, debugging, and developing modular computational solutions. Later, Weller *et al.* [25] further refined this perspective by creating a framework for computational thinking practices in introductory physics. This framework included general descriptions of the computational thinking practices of physics students, as well as descriptions of how students engaged in them. For instance, the practice "adding complexity to a model" describes how students would, when prompted, add a damping force to a simulation of a spring. Other practices included "debugging" and "analyzing data," the descriptions of which we will later leverage in understanding the differences between computational physics practices at the introductory and more advanced levels.

Finally, there is a strand of research on *scientific practices*. This strand is highly relevant to our work, as it describes how students learn to use different computational practices to do the work of professional physicists. Irving *et al.* [18], for instance, describe a curriculum with projects, assessments, and group work that are all designed to support students in learning physics practices, where computation is a central component. Odden *et al.* provide another example of practice-focused instruction, describing the use of computational essays in introductory physics to scaffold students in engaging in researchlike projects [10] and take up epistemic agency within their courses [26].

Although this literature is robust, according to the overview by Odden and Caballero [7], there exists limited research on computational physics education beyond the introductory level. The recent resource letter for computational physics [27] also notes a significant opportunity for future work on integrating computation within upper-level courses. It includes only four articles in the category, none of which include programming beyond an introductory level [28–31]. We therefore lack research on how students take the modeling, computational thinking, or research practices learned in their introductory courses and develop them into skills that can be effectively used in physics research or industry.

Of the literature that exists on more advanced programming in physics education, most has focused on the design of project-based computational physics courses. For instance, Burke and Atherton [9] created an epistemic model for expert problem solving in computational physics and described how they built an undergraduate computational physics course to teach these practices. Their model describes the interplay of computational physics practices like numerical analysis, physical transcription, and visualization, as well as implementing, running, and testing code. This model highlights how physics and coding are not done separately, but instead inform one another. One example of this interplay was the importance of visualization for planning and implementing effective programs.

Building on this work, Phillips *et al.* [32], further developed both the epistemic model and the course designed by Burke and Atherton [9]. Their revised epistemic model goes even further in emphasizing how frequently and arbitrarily scientists move between different practices within the domain of computational modeling such as planning, interpreting, debugging, and visualizing. Their epistemic model also describes the computational modeling process in more detail, including its objectives, resources that scientists draw on, and the products made during the modeling process. Although we are interested in computational modeling in this paper, we will use a less detailed framing than Phillips *et al.*, instead taking a broader view (described in the next section) which gives more focus to programming and collaboration.

Graves and Light [33] drew more directly on the expert perspective, surveying expert computational physicists on what computational physics knowledge they required for their research and whether these skills should be learned before or during research. The surveyed experts noted the importance of learning programming basics beforehand, in addition to the importance of communication skills for research. Some experts noted that they were self-taught in professional computational physics but did not see this as the best ways for future students to learn. Outside of PER, we find more research on the practices of professional computational scientists. In a series of "ten simple rules" papers [34–36], various practitioners' advice is given on how to efficiently write research code—that is, code for research that goes beyond simple scripts but is at a lower level of complexity than enterprise software. These papers highlight practices important for robust and reproducible computational research, such as testing, documentation, and version control; as well as practices that enhance efficiency, such as prototyping, refactoring, not optimizing too early, and working from established tools. The papers do not, however, specifically address the needs for computational physics, nor how these practices should be taught. Hamilton [37] provides such a perspective, demonstrating the design of learning scaffolds and a progression of numerical methods across three computational physics courses, from high school to entry-level graduate level. The final course was mainly project-based in its assessment and covered a wide range of numerical methods.

Looking across these different bodies of work, there is a clear gap in the literature on computational physics education beyond the introductory level. More specifically, there is limited research on both the skills that physics students need to develop while preparing for computational physics research and the ways in which students can develop these skills. Addressing this gap is important, as the field of PER has a responsibility to help build a workforce of physics researchers who can navigate an increasingly computational world [6].

## III. THEORETICAL FRAMEWORK: PHYSICS COMPUTATIONAL LITERACY

To address this gap in the literature, we need a framework to describe these various aspects of computation in physics that cuts across these different strands of research. Physics computational literacy, as Odden and Zwickl [38] and Odden and Caballero [7] argue, is a productive framework for informing such PER work, as it captures both the disciplinary and multifaceted nature of computation in physics while incorporating key elements of other frameworks like computational modeling, practices, and computational thinking.

*Physics computational literacy* is a form of disciplinary computational literacy proposed and used by Odden *et al.* [12,26,39]. Physics computational literacy (PCL) builds on the broader theory of *computational literacy* (CL) coined by diSessa [40,41]. In this view, computation is a literacy similar to textual and mathematical literacy, and like those other literacies, it unlocks new ways of thinking, learning, and communicating. In this disciplinary view of CL, we take into account the specific goals and practices of physics in order to study computation as an integral component of the discipline of physics [38,42,43]. In other words, physicists have their own specific sets of goals when using computation, which may differ from those of other disciplines: they model phenomena, analyze data, run experiments, and more. They also have their own ways of achieving these goals, specific to the kinds of problems and analyses common to physics. For modeling a many-body quantum system, for instance, they might use Monte Carlo methods, Python code libraries like NumPy and SciPym, and techniques like parallelization to speed up large simulations. They then review and present their models in physics research articles or conference presentations.

Computational literacy (both generic and disciplinary) is split into three pillars: material, cognitive, and social. We will use these pillars to structure our discussion of how students use computation throughout the rest of this article. However, although we will sort student practices into the three pillars, their general inclusion and discussion will be more important than their exact placement, as the primary purpose of these pillars is to identify and highlight important aspects of computational physics. Thus, these pillars constitute a complete literacy that is often hard to disentangle.

The *material* pillar relates to how students use and understand code syntax, structures, and coding tools, independent of domain. Students with a foundational level of literacy in the material pillar are familiar with some fundamental language features, like variables, conditionals, and loops [12]. We also place more advanced coding practices in this pillar, like writing modular code, using object orientation, as well as familiarity with code editors and documentation.

The *cognitive* pillar relates to applying code to domain-specific problems, in our case using computation to solve physics problems, analyze data in physics, build and run physics models, or interface with equipment. A student with a foundational level of literacy in the cognitive pillar is able to computationally model systems from an introductory physics course and interpret the results, like simulating projectile motion with air resistance [12]. When students translate physics systems, concepts, and algorithms into code, they draw on the cognitive pillar. Practices such as creating and manipulating visualizations using computation also fall under this pillar.

The *social* pillar relates to how students collaborate on code and communicate about that code. Students with a foundational literacy in the social pillar can explain a simple piece of code to others, as well as plan and discuss simple code structures. They can also consider how others will understand their code and take this into account when writing or talking about it [12]. Literacy in the material pillar is somewhat of a prerequisite to literacy in the social pillar, since understanding syntax and code structures is important to reasoning about code in this way. The social pillar also relates to communicating computational results using visualization, documentation, or polished code examples.

Most activities where students use computation will include elements from several of these pillars. When students discuss how to model a rocket, for instance, they might draw on the cognitive pillar to translate the relevant physics into a plausible numerical solution, the material pillar to recognize and understand the relevant code structures needed to address the problem, like variables and loops, and the social pillar to effectively communicate these various aspects to a grader or peer. The value of the pillars, then, is not to sort activities but to illuminate different aspects of what the students do when working with computation in physics.

The theory of physics computational literacy incorporates many of the elements that have been classified as CT in the PER literature. Computational thinking, as defined by Wing [21], is similar to how we have described the cognitive pillar of computational literacy, as both mainly relate to applying computation to solving problems. PCL also incorporates elements from both the practice perspective, highlighting the different skills and practices students need in order to become computationally literate in physics, and the computational modeling perspective, which cuts across the three pillars [7,12].

In addition to leveraging the three pillars of PCL to identify key elements of computational literacy above the introductory level, in this study, we also focus on *where* and *how* students acquire these elements. To do so, we are drawing on a second body of work, the novice-expert framework [44]. This framework takes a holistic perspective on the process of learning; rather than framing learning as a process of acquiring individual skills or concepts, it frames it as a process of gradually taking up the practices, attitudes, and habits of mind of a professional or expert in a particular domain or profession. Prior physics education research on problem solving has used this framework to examine how physics students transition from novice to expert problem solvers [45,46], showing how students' views of and fluency with the problem-solving process become increasingly sophisticated. In line with this body of work, in this study, we will draw on this framework to identify how students develop from novice to journeyman computational physicists, thereby situating this work within the field's understanding of how students acquire expertise in physics.

Taking inspiration from Dreyfus and Dreyfus [47], we recognize competence at the novice level as mainly consisting of a set of foundational skills along with a set of somewhat rigid rules and procedures for applying them to simple, known problems. For instance, a novice computational physicist might know how to make a function, comment a line of code, and create a plot or an integration loop. They can apply these skills to simple problems, like those on an assignment or exam, but do not yet know when or how to apply these elements to unstructured problems like those encountered in research.

Competence at the journeyman level consists of higher-level skills along with *situational knowledge*, that is, knowledge that is harder to formalize, but is learned through experience. A journeyman computational physicist may know when to optimize code, the different ways to document their code and their realms of applicability, and when to organize code into functions. Students at this level are able to discern many of the most important elements of a problem or piece of code and can more easily reason about more complex problems. They are purposeful in their choice of strategies and tools, and are more fluid in their work, no longer needing to deliberate on basic elements.

At the expert level, the previous elements of competence are developed further, allowing a much higher level of fluidity, complexity, and abstraction. Although we see no evidence of students gaining this level of intuition and mastery in our data, we use it to signify a goal that students approach when engaging in professional research.

With our theoretical frameworks defined, we formulate the following research questions:

1. *How do computational physics students develop their computational literacy to the journeyman level?*
2. *What are the essential components of physics computational literacy at the journeyman level?*

## IV. CONTEXT AND METHODS

### A. Context: The University of Oslo computational physics program

As we want to study how students acquire expertise in computational physics, especially at the journeyman level, we need a population of physics students with experience using computation in physics throughout both their bachelor's and master's studies. We therefore study students enrolled in a computational physics master's program at the University of Oslo. Being a Norwegian university, most students in the physics degree program have taken introductory physics in high school, and, since 2020, most newly enrolled students also have some experience with programming from high school [48]. In Norway, a bachelor's degree in physics is a three-year degree, with three courses per semester for a full-time student. Around half of the students who finish a bachelor's degree then go on to take a two-year master's degree, culminating in a published master's thesis. A much smaller number then go on to take a three-year Ph.D.

The physics curriculum at the University of Oslo has for many years been designed with computation as an integral component, and this curricular approach has been the object of several previous studies [10,12,26]. The computational elements of the curriculum are structured as follows: at the beginning of their bachelor's studies, students are taught the basics of calculus, mechanics, programming, and computational modeling. In their very first semester, they take a course in programming for scientific applications and a mechanics course with an emphasis on computational modeling. This first semester gives the students a good baseline in Python programming, which means that later courses can assume a certain degree of competence with computation.

After the first semester, students go on to use computation in most of their physics courses, where computational methods are incorporated to varying degrees. In their fourth-semester advanced lab course, for instance, students collect, process, and visualize data with the help of programming. In both their third-semester electromagnetism course and fourth-semester waves and oscillations course, students solve computational problems and write *computational essays* [12,26], open-ended assignments where students model and explore a physics phenomenon related to the course syllabus, and then write a report incorporating their code and narrative in a computational notebook environment [49].

Later in the bachelor's degree, and at the beginning of the master's degree, some physics courses have an even bigger emphasis on computation. This emphasis begins in the third semester, when many students take a course in astronomy that has larger and more complex programming tasks than previous courses, with the option of completing a large computational project instead of a final exam. From the fifth semester, students can take a dedicated computational physics course in which they learn a new programming language, C++, and complete five large computational projects. Both of these project-based courses challenge students by requiring them to implement an array of numerical methods, like the Runge-Kutta method or Markov chain Monte Carlo, or by requiring students to account for multiple complicating factors like multiple particles, multiple forces, and parallelization or other optimizations. Results from these models are also typically visualized in several different ways, as different methods and parameters for these methods are tested and compared. Students commonly work in groups on these projects, adding the challenge of collaborating and distributing tasks among group members.

The University of Oslo also offers a project-based physics course in machine learning, commonly taken in the fifth semester of the physics bachelor's or first semester of a physics master's degree, where students implement machine learning models like neural networks and logistic regression from scratch, and then use them to model several types of data interchangeably. Beyond these, there are even more advanced courses in computational physics, machine learning, and high-performance computing, which students can take as capstone courses to their studies.

For students doing a computational physics master's degree at the University of Oslo, then, the educational component consists of three years of bachelor's courses and one year of master's courses, after which the master's students write their thesis. For computational physics master's theses, the potential topics are quite varied, with some students running simulations, some working with experimental data, and others focusing more on theory or machine learning. Some students join existing research groups at this stage, while others work on individual projects where they rely more on their supervisor and student cohort for help.

### B. Data collection

In order to answer the research questions, we need a research method that will let us probe both how students from the sampled population use programming in their studies and how they learned these skills. We have chosen a thematic analysis of retrospective semistructured interviews as our method.

The study population consisted of 13 computational physics master's students, 12 male and 1 female. All participants were recruited at the students' joint offices by the first author and were incentivized with a gift card for their participation. Three of the students were in their first semester of the master's program, two were in their second, one was in their third, and seven had just delivered their theses.

The interviews were held in-person, in English or Norwegian, and lasted around an hour. They followed an interview protocol designed to probe each of the three pillars of the students' physics computational literacy, as well as the students' experiences learning programming. It is worth noting that the first author, who conducted the interviews, had themselves completed a computational physics master's degree in this same program a few years prior, and experience from this program guided the development of the protocol. Although the interviews were structured by a protocol, they were in practice flexible and somewhat open-ended because our research is exploratory, and we did not yet know which factors were most important to students' development of PCL beyond the novice level.

Interviews generally started with questions regarding the students' thesis work, including the computational methods they used. This line of questioning served as a warm-up. In cases where students had finished their thesis, the interviewer would skim the thesis beforehand and prepare questions specifically probing relevant computational work, like asking how they learned to handle specific types of data or why they chose to study a particular method. Based on these initial discussions, the interviewer next focused on the material pillar with questions about the students' choice of programming toolkit, strategies, and opinions. Mixed in with these questions were follow-ups on the students' approach to computational modeling or data analysis, relevant to the cognitive pillar. From there, the interview generally shifted to questions meant to elicit elements of the social pillar, such as how students collaborated on code or discussed and presented code. At the end of the interview, students were asked about where and how they developed the skills and opinions they had discussed so far, with a specific focus on the role of their coursework in this development. The complete interview protocol is included in Supplemental Material to this article [50].

All interviews were recorded using both an audio recorder and a separate screen recording of the student's laptop, where they showed their code, in order to document and discuss code strategy, complexity, and style. Interviews were then transcribed and anonymized, with names replaced with pseudonyms. Interview excerpts presented in this article have been translated from Norwegian into English where necessary.

Whenever possible, we also collected the students' theses, either finished or in progress, and received permission to view code they had written for their theses and coursework, which most students had saved on GitHub.

### C. Analysis

Interviews were analyzed using a six-stage thematic analysis methodology, as described by Braun and Clarke [51]. This analysis proceeded as follows: First, while transcribing the interviews, the first author noted down quotes deemed relevant to the analysis and grouped them into themes; this constituted the first two stages of "familiarizing yourself with your data" and "generating initial codes." Based on these themes and the research questions, the first author constructed an initial coding scheme and used it to code 5 out of the 13 interviews, constituting the third stage, "searching for themes." Thereafter, all authors reviewed and discussed the themes constructed. Based on this review, the first author revised the coding scheme and applied it to the data a second time. During this round of coding, the first author coded approximately 10% of the data and then discussed the revised codes and choices with the second author, which resulted in minor modifications. This constituted the fourth stage, "reviewing themes." The final coding scheme included codes for the elements of PCL, which are shown in Figs. 1–4, as well as codes for where students learned and developed these elements. This coding scheme was then applied to the remaining 90% of the data by the first author. Finally, the definitions of the themes were refined during the writing of this article, in accordance with the final two stages "defining and naming themes" and "producing the report."

In addition to this interview analysis, we triangulated themes from the interviews with analysis of student theses and code repositories. These artifacts let us check several aspects of the students' self-reported PCL, such as whether the coding practices and preferences students described in fact matched their actual work, or how their early work differed from their newer work. This check focused mostly on the structure, readability, and size of their code, as well as the languages and code libraries they used. As the selection of courses and the set of thesis topics were quite varied, we did not formalize and structure the analysis of these artifacts; it therefore acts as supplementary analysis to the main findings from the interviews.

## V. RESULTS I: DEVELOPMENT OF PHYSICS COMPUTATIONAL LITERACY AT THE JOURNEYMAN LEVEL

We begin with research question 1, *How do computational physics students develop their computational literacy to the journeyman level?*

Based on the interviews with computational physics master's students, we identified two distinct transitions in students' development of PCL beyond the novice level, as well as important factors supporting the ongoing development of their computational physics expertise. Students commonly described the first transition in their PCL when they encountered a project-based computational course in physics. We will denote this change as an initial transition from the novice to the journeyman level. The second transition happened when students participated in professional, collaborative computational research, which several students did during their thesis projects. We will denote this stage as the students' preliminary development of research competence, through which they further progress into a journeyman level of expertise. We found this change to be smaller and less explicit than the first, which is to be expected due to the fact that master's projects often have a limited scope and are mainly done independently. In addition to these two transitions, we found that self-directed learning and peer interaction supported the ongoing development of PCL both at the novice and journeyman levels, as students often relied on sources outside their lectures, instructors, and course curricula to improve their PCL.

### A. Transitioning from the novice to the journeyman level of computational physics expertise

Interviewed students clearly differentiated their competence within computational physics before and after they took advanced physics courses that featured complex, project-based computational work. Across the interviewed students, there was a general consensus that introductory courses provided a solid foundation for their physics computational literacy, but that the project-based courses they had taken from their third semester onward seemed to provide skills more applicable to the research work that they would eventually do in their master's theses. For example, one student stated it simply as follows:

Nils: You don't really learn to code in [an] intro course.

Students often mentioned their first project-based course, a third-semester astrophysics course called AST2000, as an important turning point in their formal training. This course acted as something of a "crucible course" in computational physics:

Jake: The intro to scientific programming course was nice, but it was AST2000 and the computational physics course that really set the programming.

A significant part of this development stemmed from the fact that the frequent computational assignments in these courses gave students experience in writing code. However, several other aspects of these courses were equally important in supporting this transition in expertise. First, the social elements of these project-based courses were crucial, as they provided students with opportunities to work with and learn from their peers:

Casey: It's a very good argument for group projects too, I think. Because you … you see how others code, and then you get a lot of inspiration, and then you also get like, I think I'm doing this in a better way. And then you become more aware that maybe it was a good way I did it. Comparing yourself to others and how they do it can be quite beneficial, within programming.

Next, large computational projects were cited as a critical forum for developing computational problem-solving skills, in that they forced students to learn how to focus on the most important aspects of a problem and abstract away the rest:

Jake: After COMP-FYS, I learned that, okay, now I have such large projects, which require so much code and so many files, that I can't have an overview in my head of what's going on here.

In addition to this increased competence, students experienced greater epistemic agency and creativity in these courses than in foundational computational physics courses. This sense of creativity and ownership foreshadowed the open-ended master's projects they would take on later in their studies:

Jason: Since it's something you kind of create with someone else and then hand in, I feel like I got more ownership of it.

Pedagogically, students contrasted these project-based courses with lecture-based courses, arguing that they learned significantly more from the former than the latter. Regarding lectures, they often spoke critically of the computational learning in those settings:

Ivar: I haven't had so many lecturers who have shown programming that I've felt I've benefited from much, really.

### B. Progressing into the journeyman level of PCL by developing research competence

The second clear transition we identified in the students' physics computational literacy occurred during the writing of their master's theses. Computational physics master's students work on their thesis for a full academic year and thus get extensive experience working on a single topic. During this year, some students develop a level of specialization and professionalization that we characterize as a part of the journeyman level of expertise. For instance, some students described a greater fluidity with the methods they specialized in:

Jonas: You learned the theory [in the courses] but you didn't implement it 10 times so it's kind of hard to remember how to implement it when you need to use it, put that theory to practice. But now I've done it a lot of times so now it's kind of in my muscle memory.

Some students described how they became more purposeful and efficient in the way they write code, either from working with professionals or from the experiences they gained working on a larger open-ended project:

Levi: I learned to implement OOP [object-oriented programming], but it was only in the master's that I learned why it's actually useful.

Students also engaged with advanced, specialized topics during their thesis work. These topics ranged from particle physics and Feynman diagrams to custom machine learning models and complex data pipelines. Learning these advanced topics added to the students' growing base of expertise:

Alexander: [I learned] the entire stack and the custom data set transformation from the real data. And that's nontrivial. Actually going to the lab and doing an experiment until having some pruned label data which can be used for machine learning is quite a journey.

Several students highlighted the important role their supervisors played in helping them learn these topics, both through their direct guidance and through helpful code examples. In this way, supervisors played a clear role in helping students continue their development as journeyman computational physicists:

Casey: I was given a lot of guidance. I got good help. I started by getting some examples from my supervisor.

Not all students experienced this development of their PCL during their thesis work, however, as their projects did not entail large computational challenges. Some students focused more on theoretical aspects, while others worked mostly on their own, not being notably challenged on their approach to coding by their supervisor, group, or project.

### C. Developing PCL through self-directed learning and peer interaction

In addition to these two distinct transitions, students described a continual development of their physics computational literacy through consistent engagement with several other sources: self-study of online resources, AI, peers, and on-the-job training.

Outside of courses, students cited online sources as the second most useful resource for learning the coding practices necessary for computational physics research:

Bruce: Courses really fail at teaching you how to write good code, but there are many tutorials on the internet, and you can try for yourself.

These resources included online forums like Stack Overflow, tutorial videos on platforms like YouTube, example code on GitHub, and other sources found through search engines like Google:

Ivar: When I started googling, when I started writing code, I used Stack Exchange quite a bit. So there are a few tips here and there, that just shape my approach.

Although these interviews were conducted approximately a year after the initial release of ChatGPT, students also found AI tools a useful aid in learning how to perform different computational physics tasks and analyses:

Casey: I spent a lot of time on it, talking to friends and stuff, but also I've actually talked quite a bit with ChatGPT, like, I have this system, this scheme, how do you recommend, or what would be a natural way to structure this code?

As in professional research settings, master's students had access to a community of peers where they both contributed and benefited from the shared knowledge of the community. Students noted that this feedback and critique from fellow students was helpful for improving their coding practices:

Steven: Others criticized my code, and I tried to fix that.

Students also shared tips with each other on computational tools and techniques they could use:

Jake: I was indoctrinated into VS Code by some friends in the study program.

Students who had worked in groups were very positive about group work, especially as a way to learn from others and gain confidence:

Casey: When you then get to be the one who maybe knows more sometimes, it's a bit nice. Not only is it good for self-confidence and stuff, but then you actually reflect on what you prefer yourself. And what … You turn on a more critical sense.

A number of students also cited "on-the-job" training as useful for their development of computational literacy, both within and outside the research world:

Jake: But it wasn't until last summer that I worked at astrophysics, I had a summer job there, they were like … "Create virtual environments."

## VI. RESULTS II: COMPONENTS OF PHYSICS COMPUTATIONAL LITERACY AT THE JOURNEYMAN LEVEL

Having identified key transitions in the progression from novice to journeyman computational physicist, we now address research question 2, "What are the essential components of physics computational literacy at the journeyman level?" To address this question, we have identified specific elements of PCL that the sampled students acquired through these different stages, across the material, cognitive, and social pillars of physics computational literacy. Figure 1 provides an overview of these components, which we will unpack in the following sections.

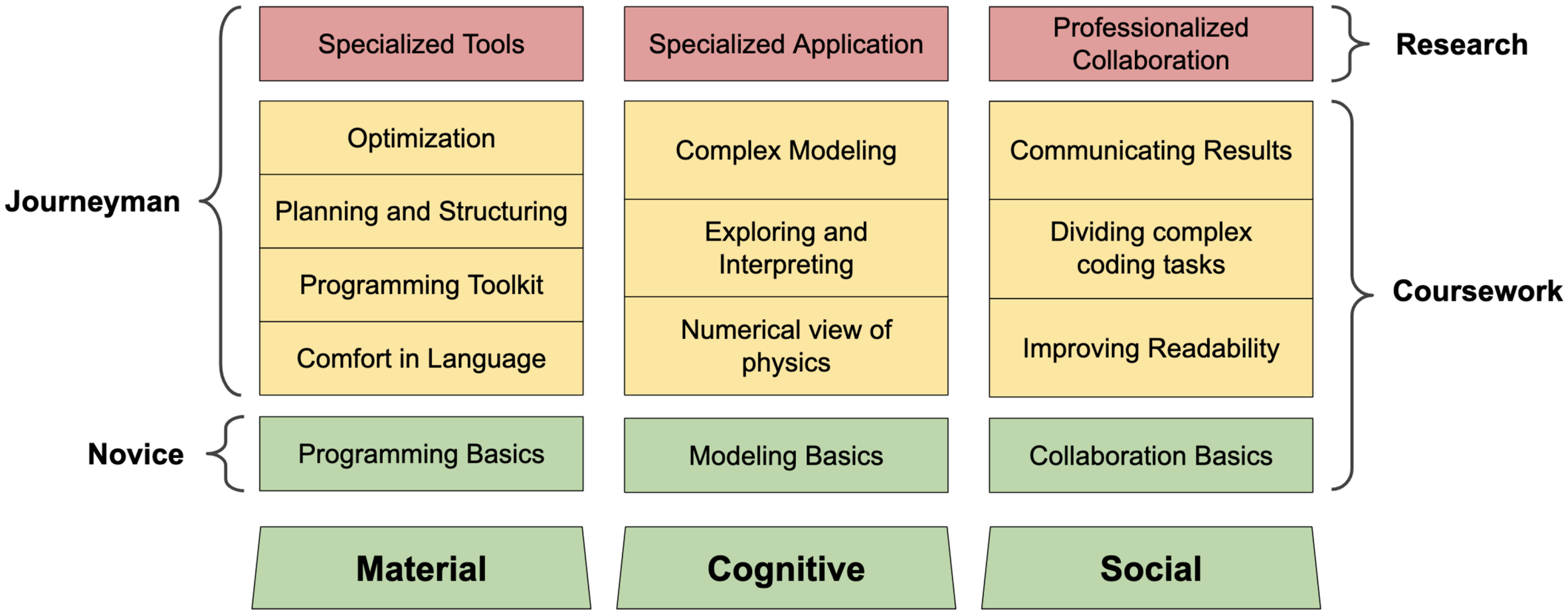

FIG. 1. The three pillars of physics computational literacy at the novice and journeyman level. Elements at the novice level act as a foundation for doing computational physics. Elements at the journeyman level are varied and complex and are learned from experience with large, collaborative projects. Elements at the top of the journeyman level describe the specialized computational skills some students develop while doing advanced research.

Across the three pillars, we see that students retain novice-level skills as a foundation to their PCL, such as writing and running code, implementing simple methods like integration loops and plotting the results, as well as working on and talking about code with others at a basic level. Journeyman-level elements of PCL build on this foundation. Starting with the *material* pillar, we see that students at a journeyman level are proficient and confident in a programming language, as well as a wider set of programming tools such as code editors, version control, and code libraries for tasks like plotting and machine learning. Students also learn ways of making code run faster, as well as the planning and structuring necessary for writing larger, more complex code efficiently. Related to the *cognitive* pillar, we see that students at the journeyman level not only learn more complex numerical methods and how to apply them to more complex problems but also develop a numerical view of physics, wherein the students naturally look for ways to solve problems with tools like discretization, data analysis, or iteration. Supporting these elements is the practice of exploring and interpreting these numerical solutions, either by making visualizations or by interrogating data. Within the *social* pillar, we found that students at the journeyman level prioritized readability of code when they thought it would be used later or by others. Students also had several opinions and practices for making their code readable. At this level, collaboration became more efficient, as students were better able to coordinate the splitting and combining of coding work. Students also developed skills for communicating their results through visualizations, code examples, and code explanations.

Finally, across all three pillars, we have identified a level of research competence that is different from what was gained through coursework, constituting a set of specialized skills that some students gained from taking part in more professional research environments and projects. These skills include knowing how to use specialized tools, such as frameworks for working with particle collider data and symbolic algebra for many-body quantum physics; specialized applications of computation, such as developing and evaluating new numerical methods; and advanced collaboration skills, like writing code for a professional research group within an established code framework.

### A. Material PCL at the journeyman level

We begin by describing identified aspects of the students' journeyman-level PCL within the material pillar. This pillar describes the students' mastery of programming syntax, tools, and strategy. Competence in this pillar is vital for professional computational physics research, as writing, running, and improving code are core parts of building computational models and research tools. Figure 2 provides a summary of this section's results.

#### *1. Programming basics*

The foundation of the material pillar is made up of basic programming skills. These include understanding and being able to use basic syntax and language features, like functions and loops, as well as being able to write and run code in a code editor. All interviewed students reported that they had learned these skills in an introductory course in programming, which they saw as helpful for their later progression:

Jake: I think what the introduction to scientific programming course really was able to do was just emphasize the importance of learning to read code. […] That really helps, because if you can read code, it is much easier to write code. Then you can also understand the code of others, so you can learn code easier.

| **Material** | |
|---|---|
| Specialized tools | Advanced programming tools that are not part of course syllabi, but that students use to aid in research (clusters, profilers, unit testing, debuggers) |
| Optimization | Knowing techniques for making code faster and when to apply them. |
| Planning and structuring | Knowing strategies for making code easy to write or extend, and opinions about these strategies (starting with a simple case, planning functions or classes, object oriented programming, functional programming) |
| Programming toolkit | Familiarity with tools like code editors, version control and AI. Knowing how and when to use different packages and languages. |
| Comfort in language | Being able to utilize advanced features in a programming language, such as classes and list comprehensions, not needing to look up basics, and being comfortable quickly learning new features and packages in the language. |
| Programming basics | Familiarity with the basic syntax and language features of a programming language. Being able to parse and debug code with rudimentary skill. |

FIG. 2. Names and descriptions of the elements of physics computational literacy in the material pillar at the novice and journeyman levels.

In addition to a handle on the fundamentals of coding, students at the novice level gain familiarity with its foundational tools, like editors, terminals, and the file system.

Andrew: Having an introductory course in programming is very good. Because when I started the bachelor's I didn't even know what a terminal was. I didn't know how to open a terminal and make it do things. I knew absolutely nothing about programming.

Although programming basics are an important and nontrivial element of PCL, we will not go into more detail on them here, as they are not a focus of our study and are described well elsewhere [12,24,25].

### *2. Comfort in language*

In computational physics research, code is the primary modality for the research work. This means that to be an effective computational researcher, one must gain a degree of comfort and fluency with this modality. The interviewed students highlighted this fluency, arguing that there is a difference between knowing the basics of coding and really knowing "how to code." More specifically, at this journeyman level, students knew enough about different language features and how they should be used that they could quickly solve familiar problems without needing to look things up. At this level, they could tell *why* a class, function, or loop should be used, in addition to the more technical details of *how* these language features work, such as variable scoping and immutability. This level of comfort coincided with the recognition of coding as a powerful and widely applicable tool:

Jake: It started out very abstract for me. Coding was like, okay, but what the heck is going on? What is a for loop? What am I doing here? To the point that it is become a very useful tool. I kind of see, what can I say, you have it like a hammer in a way. You learn to recognize what you can do with it.

Beyond comfort in solving familiar problems with familiar tools, students at this stage were also confident in their ability to learn new tools and solve new problems. This confidence came in part from their learning to use documentation (online documents that explain coding tools, features, and packages) effectively:

Christian: I feel proficient enough that if I get a problem that isn't too advanced or too technical, then it wouldn't take me a long time to just read up on the documentation, and get familiar enough with it to be able to solve it, kind of.

Students reported that they were all very comfortable using the Python programming language, which is used in introductory courses at the University of Oslo and most other courses in the physics degree program. Some students also felt very comfortable using other languages such as C++, C, and MATLAB. The students mostly used these languages when their courses or other external factors required them to, for instance, needing to implement an algorithm in C++ for the sake of speed. However, they generally saw C++ as more unwieldy than Python, and Python was typically the language students naturally gravitated toward when thinking through solutions to their computational problems:

Jonas: C++, that was a lot of fun once you learn how to write in C++. But I mean it's much harder, and the same with C.

### 3. Programming toolkit

In computational physics, one needs to know more about coding than just how to use a programming language: one also needs to know how to use an assortment of supporting tools that play an important part in developing and improving code. For the sampled students doing computational physics at the journeyman level, these tools included editors, terminals, AI, online resources, and a variety of programming languages and packages.

Code editors form the basis for code development: one needs an editor to create, edit, and save code, and in order to run code, one needs to know the correct terminal or editor commands. Becoming efficient with editors and terminals can be a major hurdle for students first learning to program, but with time and experience, fluency with these tools can greatly enhance the coding efficiency of students, as modern code editors include several helpful features. Code completion and parameter information, for instance, can give relevant insight into selections of code and even suggest how to complete a line of code once one has begun writing it. These tools are typically enabled by default, and the interviewed students described how said tools quickly became ingrained in their programming workflows. Most interviewed students used the editor VS Code:

Jason: Everyone else in the world uses VS Code, let's try that, and there is a reason they do that. Of course they were right.

Other code editing tools were used by a smaller number of interviewed students, as these tools were not taught in their courses. These tools included the debugger, which gives information about code at certain "breakpoints" while it runs, and formatters, which edit code to make it easier to read. However, even students who did not use these tools were often aware of them:

Steven: I do a lot of print debugging. I haven't used a debugger or something like that, but I have stuff installed in my VS code that will point out unused variables, usage of undeclared variables. And they really help me a lot to point out where the problem is.

AI tools like Copilot, which suggest code within the code editor, and ChatGPT, which answers questions in a chat window, were also noted as useful tools for coding more efficiently. Students used these AIs for various tasks, like quickly fixing bugs or producing code for plotting and visualization, even though these tools were relatively new at the time of the interviews (early 2023):

Ivar: I actually use it for everything. If it's just simple repetitive tasks that I can't be bothered to do manually, like setting up pretty simple algorithms. Then instead of writing it completely from scratch, maybe it's very… Say it's an old algorithm that I'm familiar with, but that I can't entirely remember of the top of my head how to set up, naturally. Then I use ChatGPT to just give me the code skeleton.

Looking up things online was also an important part of the interviewed students' workflows. Both programming forums, like Stack Overflow, and documentation websites (user manuals) played an important role not only in helping students learn programming but also in assisting students in solving programming problems, such as fixing bugs:

Steven: Oh yeah. When I'm debugging my code, I firstly try to understand what I'm doing. But when I'm writing the code, I use a lot of Stack Overflow to figure out how do I do this thing. And of course, if I am able to identify a bug in the code and don't immediately see how I would fix it, Stack Overflow is invaluable.

Students often measured their proficiency in programming in terms of their ability to efficiently solve problems using only documentation as a resource and did not particularly value remembering every detail about a language or package. However, students still found it valuable to be familiar with a range of languages and packages. That is, at the journeyman level, many students knew how to use a broader toolkit of languages and packages and were able to select and use the appropriate language and packages for the problem at hand. For the interviewed students, the typical example was the aforementioned familiarity with the language C++, which they chose to use for some high-performance-computing tasks, even though they were more comfortable using Python.

### 4. Planning and structuring

When a codebase grows large, as it often does when solving complex problems in physics research or when working in collaborations, there is a need to consider its structure in order to maintain efficiency. Accordingly, students at the journeyman level of PCL develop a mindful approach to planning and structuring larger code.

Most students began planning a computational project by spending some time understanding the problem before writing any code. Some spent time working on the problem analytically, while others asked ChatGPT for ideas on how to approach it, and still others focused on determining which pieces of code they would need to write:

Ivar: I always try to do as much as I can analytically, perhaps. So if I have a physics problem, then it would be to find out as much as I can about the problem before I start coding, no matter what. There are so many times I start writing code, and then I find out that this was actually wrong, and so I have to start removing the code again.

Once students actually began to plan the structure of their code, there was more variation in approaches, with some preferring to first plan out the complete structure of the codebase, while many others preferred to just start coding the simplest parts and expanding from there. Those who preferred planning out their code in advance sometimes represented their code structure using drawn schematics or created a skeleton code out of empty functions and classes. We note that such processes were not explicitly taught in the classes these students had taken, so their planning process was not formalized. This process of creating skeleton code was nicely illustrated by a student we have chosen to call Casey:

Casey: But at least in the computational physics course, we typically got some physics problem, and then, if we were then going to solve it, I started making the class and the class structure I thought I wanted. I first kind of just set up almost just the titles of things. I just set up, OK, I want that, my "run" should do this, I want another class that will belong to this class in this way, or inherit from this class in this way. Start by doing that kind of thing, and then I start filling it in, maybe.

However, Casey also noted that this approach of starting with a class structure is less applicable when the problem and solution are less clearly defined, as was the case in their thesis work. In these cases, the second approach was more applicable:

Casey: When I more kind of have a problem … in my master's thesis, that maybe is a bit more undefined, I don't know where I'm going or where I start, then I can't do that. Then it's much more about trying a little bit here and a little bit there, and then maybe the code comes more and more together over time. I understand more and more what I need.

Other students used this more organic approach as their go-to strategy, viewing planning a complete code structure in advance as a poor choice:

Levi: What I like to do is to run incomplete code to see where it breaks, for example. […] Because it's more palpable than just like, oh, I'm going to only start coding after I have the whole code structure in my head. Some people do that. I cannot. I cannot do that. I think it's not productive because usually it will fail in points that you don't think it will. Even if you devise your brilliant plan.

Regardless of their approach, most students preferred to implement a simple solution first and then add more complex parts to the code only after that simple solution worked:

Alfred: But I think to get a bird's eye view perhaps on what you're supposed to do, and then put up a structure, I want this and this and this. And then just get the minimal amount to run, like a test trial. If you're simulating, say, some physical system, get a simulation going for the smallest possible version of the system.

When researchers add functionality to code, the complexity and size of that code increase, and this additional complexity needs to be taken into account in the structure. Students who initially mapped out or mocked up their code structure often hoped that they could just fill in code in the functions, classes, and files they had planned out and keep everything organized along the way. Those who grew their code structures more organically instead made structural choices as they went, especially regarding object orientation and file-structuring:

Jason: If I don't have a clear picture of how I'm going to object orient it, then I don't. Then I first write everything, not functional, but with functions, basically. And then at some point I realize that, oh, this is how it could be done with classes instead.

In addition to structuring code to make it easier to keep track of and change, students also sometimes modularized their code so that they could reuse it for multiple different purposes:

George: I think you are met with the choice of making these considerations pretty often. For instance in lab courses, then … you often start writing separate code for each problem, and then you notice that you have to do this same thing many times, and then you notice that, ok, I have to structure my code so that I can use it multiple times.

All interviewed students who provided access to their master's thesis codebase were found to have used both classes and functions to organize their code. When numerical methods needed to be interchangeable or modular, these students typically wrote their code as classes. For plotting or data processing operations that were more stand-alone, these students typically wrote functions.

#### 5. Optimization

In professional computing research, computing workloads can quickly become large, and speed and efficiency in code become increasingly important. Journeyman computational physicists must begin to grapple with this need for efficiency.

The journeyman computational physicists interviewed in this study cited several different ways to speed up code. From their courses, many learned to parallelize and vectorize important calculations. Some students preferred to write code in C or C++ to optimize for speed, although many found the libraries available in Python sufficient for making efficient calculations.

When discussing these considerations, the students often noted the need to ensure that the code actually works first; only then should its optimization become a consideration:

Nils: The first and biggest hurdle is obviously the code is able to solve the problem. And then you've got other things, for instance, like, okay, well, how effective is it? Right? So how quick is it? Has it been written in a way which does as few flops as possible? Or are you kind of doing lots of extra work, which you wouldn't have to do?

Students also noted that speed was important only if the code takes a long time to run. In other cases, it can be worth writing slower code, which is easier to understand:

Alfred: If the program is running a simulation, and the simulation is heavy, then efficiency is very good and very important. But if it's a code which doesn't really matter if it takes 10 milliseconds or half a second, then I think it's better to go with the more readable code.

These considerations of when in the coding process to optimize code, and for which problems optimization is appropriate, demonstrate the intentionality and context-dependent strategies of journeyman-level PCL.

#### 6. Specialized tools

Computational physics research often relies on specialized tools built for experts, which require more setup and training than the more general tools used for learning or generic computational work.

Some interviewed students had learned to use such specialized programming tools during their master's theses, which were not taught or expected in any other part of their studies. We place these tools higher up in the journeyman level of PCL, as we view specialization in a specific area of computational physics as a more advanced aspect of gaining computational physics expertise.

Within the set of surveyed students, some learned how to use specific programs or frameworks. For instance, one student learned a framework for handling collider data, while another learned to use a symbolic algebra system for doing many-body quantum physics. Others specialized in more generally applicable tools, like machine learning frameworks, profilers that find which parts of the code are slowing down execution, and tools for running code remotely or on a cluster:

Casey: We [the research group] make a bunch of jobs, because we process a lot of data, so you would prefer that it's not run locally. So you want to set it up on remote machines.

Some students did not engage with more complex tools but instead developed further expertise with tools they had already used during their courses:

Jonas: I have now a black belt in pandas, because I used it like two hours every day. I know all the functions, and yeah. I learned to be very comfortable with setting up a framework for just processing data and creating a model, and training the model, and testing it. I got very comfortable with doing that because I did it so many times.

These specialized skills are not necessarily harder to learn than the skills that came before, but rather reflect the needs of research that go beyond what is typically acquired from coursework. Learning to use such tools is an important step toward developing computational physics expertise and preparing students for more advanced work in the field.

### B. Cognitive PCL at the journeyman level

The cognitive pillar of PCL focuses on the ability to apply skills and techniques from the material pillar to physics tasks and problems. At the foundation of the cognitive pillar are modeling basics, that is, the ability to implement simple numerical methods to simulate the behavior of physics systems and visualize their results [7,12]. At the journeyman level of PCL competence, we find that students develop the ability to explore, interpret, and reason about numerical data, as well as implement more advanced simulation and data analysis techniques on increasingly complex physics phenomena. Figure 3 provides a summary of these results.

#### 1. Modeling basics

Computational modeling is a multifaceted practice central to computational physics research, since models give researchers the ability to probe systems that are often experimentally inaccessible. Although it can take many years to gain expertise in computational modeling, students can learn fundamental modeling techniques fairly early in their physics careers, such as making sensible assumptions, discretizing analytical expressions, reading or generating data, solving equations iteratively, and interpreting and visualizing results. The foundation of the cognitive pillar is made up of these basic modeling skills.

| Cognitive | |
|---|---|
| Specialized application | Applying computation to a problem with the complexity and open-endedness of real research. |
| Complex modeling | Implementing advanced numerical methods, including machine learning. Handling a lot of data to simulate or process. |
| Exploring and interpreting | Exploring and interpreting numerical data, often before or after modeling is done. |
| Numerical view of physics | Recognizing the strengths and limitations of numerical solutions and representations. |
| Modeling basics | Implementing simple numerical methods, as well as related processing and visualization |

FIG. 3. Names and descriptions of the elements of physics computational literacy in the cognitive pillar at the novice and journeyman levels.

Although we call these skills "basic" in the context of computational physics master's students, they are not easy to learn when starting out. For example, already in the first semester, students at the University of Oslo implement integrators that can be used to simulate complex physical systems, such as projectile motion with air resistance, and many students find these types of techniques challenging to understand:

Jake: So what I found hard in the start of the intro mechanics course was when we started with the Euler-method and things like that. […] I found it hard in the beginning, because it was like the first time I had to think in algorithms, but you quickly get used to it.

These basics act as a foundation for later development, in that part of reaching a journeyman level of competence involves internalizing several of these steps, such as which physical assumptions to make, and how to reason about discrete data and iterative solvers. Once these aspects are internalized, it becomes easier to reason about more complex systems and methods.

We will not go into more detail on these basics, as the focus of our research and data is on what happens when physics students progress beyond this level. However, there is substantial research on teaching computational modeling in introductory physics [7,12,18,19,24,25].

### 2. Numerical view of physics

An important element of computational physics research is developing an intuition for how to approach questions or problems numerically—that is, which aspects of those questions or problems, as well as the relevant physics theory, lend themselves to numerical representation. The journeyman computational physicists we interviewed described this kind of numerical view of physics. When these students encountered a physics problem, they often framed it numerically, for example, immediately thinking of ways to discretize the problem and make it suitable for a numerical method. Numerical solutions are not necessarily easy to find and implement, however, as they require an understanding of the relevant physics, discretization, and different possible solution methods:

Ivar: You have to think about physics a little differently. Because, in addition to understanding what's happening, you have to understand how you can discretize it on a computer. And then in a way you have to have a slightly different understanding, maybe, of how things work, so you can get it into a computer. I notice it very specifically with the students I have [as a learning assistant] in electromagnetism. That many people can sort of understand the electric field, but they just don't understand how you can have the electric field on a computer, and how you can have a grid, and how it's the same as the field in space.

Several of the interviewed students, in fact, preferred numerical solutions over analytical solutions, which is perhaps natural given that they chose to specialize in computational physics. Although these students often found numerical solutions easier to understand, most did not see them as inherently simpler than analytical physics, but rather as an alternative perspective and set of methods with their own affordances and realms of applicability:

Steven: It has given me a different mathematics understanding. Because there's different mathematical tools required for numerical programming than for analytic problem solving. So like I might not be as strong on like different techniques of solving differential equations analytically. But there's a lot of considerations, like mathematical considerations, going into solving differential equations numerically.

### *3. Exploring and interpreting*

Computational physics research is often exploratory. As noted by Phillips *et al.* [32], the products of the modeling process often inform subsequent exploration. For students at the journeyman level of PCL, visualization is a key facilitator for this kind of exploration. During their coursework, students had frequent practice at generating plots, and these types of visualizations played an important part in helping the students make sure their work was correct (in combination with code debugging):

Levi: Generating a plot, it's better than like… I mean, it's more convincing than drawing some graph and it's like, oh, this is what I think is going to happen based on this equation. It's like, no, no, run the simulation, generate the points, fit the thing, see if it's actually true. So it's more convincing. And for understanding, it forces you to break the problem into small bits that are digestible.

Students also described how visualizations helped them to gain a better understanding of the different ways their algorithms worked:

Alexander: If you develop the code and you say, well this graph is made by doing this and this calculation and this is the results, then you could start better questioning why the graph is the way that it is, other than just observing that is the graph.

As the datasets students work with become more complex, students must also develop more sophisticated approaches to data analysis and exploration, such as exploring data using statistical properties or accessing small parts of multidimensional data at a time to make sense of it:

Casey: For me it's important to understand what kind of data we are working with. What does this data we have mean? So sitting and actually programming and working with the data, I think that is important. To get an understanding of what the data is. I don't necessarily get that from my plots.

### *4. Complex modeling*

When students transition into professional computational physics research, the models and phenomena they study often become increasingly complex. An important part of this transition is learning about and from an array of different methods and systems. Students in the study population had researched systems as varied as the buckling beam, the Penning trap, the Ising model, and the double-slit experiment, each of which required its own set of numerical methods and considerations.

When the methods and systems become more complex, both the code and the physics become harder to reason about:

Ivar: Often I first look at the algorithm and try to find out if it's my code that is just wrong. And if I read up and get a thorough understanding of the algorithm and the code, […] then I have to go back and look at the physics. Is that where I'm wrong? Is it the initial conditions or something like that I've done wrong?

Students particularly highlighted machine learning methods as useful for modeling more complex phenomena and found these techniques equally applicable for solving real-world problems:

Jonas: So machine learning was kind of always the plan and also it kind of had a direct real-world application. So that's why it was more interesting to me than just some experiment on simulated data.

Learning this gallery of increasingly complex methods sets students up to do more authentic, self-directed computational research.

### *5. Specialized applications*

There is more to computational physics research than just implementing complex analysis methods. In many cases, a large part of the work is finding out what to study and which methods to use. When projects are more open-ended and specialized, as in master's projects, students must take responsibility for more aspects of the project than they would in structured environments like course assignments. These can include drawing on the literature and their own expertise to determine which model to implement, as well as assessing the objectives of the modeling process and whether they have been met [32]. Adding to these challenges, many students find that the specialized or novel methods necessary for their projects are often not neatly described in a textbook or tutorial, and that they therefore need to figure out how they work on their own.

For their master's projects, the students we interviewed sometimes studied new physics, numerical methods, and

| Social | |
|---|---|
| Professionalized collaboration | Collaborating with experts, building on the work of others in a structured way. |
| Communicating results | Communicating non-trivial results using visualizations, figures and descriptions of how the code works. |
| Dividing complex coding tasks | Segmenting complex coding tasks in a productive way. |
| Improving readability | Knowing how and when to make code readable, as well as how and when to manage readability when collaborating. |
| Collaboration basics | Explaining and commenting simple code and numerical results. |

FIG. 4. Names and descriptions of the elements of physics computational literacy in the social pillar at the novice and journeyman levels.

ways of using data to solve problems. Some students, for instance, tried out new methods for simulating many-body quantum systems, or for identifying new physics in collider data. In these projects, it was often up to the students to figure out what they were looking for and how to approach it:

Nils: And I remember asking myself, well, okay, should I do one model per variant, or should I use one model to look for many variants? And when I did some tests, I found out that using one model to look at many variants was actually the best.

We place these specialized methods learned during the master's thesis work at the top of the journeyman level of cognitive PCL, as they are often more complex than methods covered during coursework, placing them closer to professional research. These methods also require a level of independence beyond course assignments or projects. We see this specialization and independence as development toward expertise in the cognitive pillar.

### C. Social PCL at the journeyman level

In addition to programming and applying computation to physics phenomena, expert researchers often collaborate with others and write code meant for others to use or adapt. When physics students collaborate on larger code projects, they encounter the inherent challenges of collaborating on code and begin to develop this competence. To navigate these challenges, students develop both general collaboration skills and learn to plan and write code in a different way than they would when working alone. In this section, we will present how the interviewed students learned to handle these challenges.

At the beginning of their studies, students collaborate on code in quite basic ways, offloading the work to one person or talking through small coding problems together. At more advanced levels, as problems and tasks become more complex, journeyman computational physics students learn collaboration skills necessary for these types of advanced problems, like dividing tasks, improving readability, and communicating results in various ways. Some students who directly collaborated with experts developed even more professionalized collaboration practices. Figure 4 provides a summary of these results.

#### *1. Collaboration basics*

When starting out, most physics students have little-to-no experience collaborating on coding assignments or projects. In reflecting on these experiences, the interviewed students mentioned that a common issue was that one student in a group ended up doing most or all of the coding, as this was easier than having everyone contribute to a single codebase. In many cases, the students we interviewed were, in fact, the ones who took on this role. At the basic level, then, students found collaboration challenging and often found these uneven workloads frustrating:

Ivar: I feel that it takes a lot longer when I explain how things should be solved. […] You have to do everything double, and if in addition to the code in itself being difficult, that it's a problem where you also have to explain the physics of the problem, then you have to explain the code, and then you have to explain how to code the physics in the problem.

For the students in this study, collaboration was rarely taught or structured, so most had to learn these skills on their own. Despite these challenges, students often tried different strategies for collaboration, such as explaining code or numerical results to each other, writing extra comments to make the code easier for others to understand, and asking others for help when they encountered problems. In this way, collaboration at this novice level of PCL prioritized learning over programming efficiency.

#### *2. Improving readability*

Computational physics research projects often span many different years, projects, group members, and collaborators. For these reasons, code in computational physics must be written to be readable and well structured in order to make it easier to understand, debug, and expand.

Naturally, then, at the journeyman level, readability of code is seen as important, especially when working or sharing code with others, as well as when codebases become large. At this level, interviewed students had developed a number of common practices to improve code readability, like using clear variable and function names, avoiding unnecessary if-statements and indentation, and commenting only when needed. Unlike novices, students at this level are able to apply these practices only when appropriate, avoiding pitfalls like overcommenting (commenting every line) or making code as concise or spacious as possible. These practices went hand-in-hand with an understanding and consideration for the level of computational literacy of collaborators:

George: Then it becomes about not commenting redundant things, like using variables that are self-explanatory so that I don't need to comment them, and so that I won't have questions about what they are. Commenting units on them, as early as possible and then not doing it from there, you should assume that the one who reads the code is mindful of that. Spacious code, and not more than three indentations.

Students at this level often leveraged code structure in order to increase readability. These considerations included both the overarching structure of the code and local structure, such as where to add spacing between lines of code or where to place comments. Interviewed students also considered placement of function and class definitions within files as important aspects of structural clarity and readability, in combination with the documentation described above:

Alfred: And good structure is then, I think, easy to navigate through. And easy to find exactly what you're looking for. To follow the flow of the program. So you might not start, or you could say, okay, it's importing this stuff. And then you have the classes and the functions and everything. But you jump straight down to where things actually happen. And then you see, okay, it's doing this and this. And then it's calling this function. And then it's easy to just go up and check, okay, this function is doing this. So it's well structured in that way. And each function and class and everything is well documented. And variable names are easy to understand.

Although these aspects have little practical impact on what the program actually does, interviewed students would sometimes go back through their code to improve these aspects in cases where the code would be useful later.

Although we have provided several examples of considerations students named when improving the readability of their code, most of these types of considerations are not necessarily easy to put into words. In line with the general characteristics of journeyman-level expertise described above, we found that students at this level did not go through a concrete checklist in their head of what choices would make their code more readable, but instead used their developed experience to intuit "good" coding choices. These choices, then, varied from person to person.

These variations in coding style sometimes led to difficulties when collaborating. Students who knew less about programming, or preferred to write code in a different way from collaborators, sometimes found it hard to contribute to groups. In certain cases, others would even rewrite their code. Learning to navigate and mediate these differences is also part of reaching the journeyman level in the social pillar, although this type of navigation often requires a more structured environment, a theme we will return to later. One student succinctly described these challenges:

Ivar: That's why I really like when everyone follows the same convention, because then you don't have that problem. But that's not something you learn in any physics course. Not everyone has that approach. It happens that I've written my part of the assignment in a class, and then they've just written with all the lines together right after each other. So now we're going to try to align these two codes, and then it becomes chaos. So often it can offer a little more challenges just because everyone hasn't been trained with the same convention.

#### *3. Dividing complex coding tasks*

Projects in computational physics research are often complex, with multiple interconnected components and several people working together. In these advanced projects, then, making one person do all the coding will not work; distributed and planned cooperation is needed.

At the journeyman level, interviewed students had learned through experience several ways to divide and structure group work. Some students took a very analytical approach, trying to find the most effective way to distribute tasks:

Nils: When I collaborate with these computational projects, I normally begin by kind of assessing what kind of role I should have. In some cases, I might be the dig deep into a problem kind of guy. So you have the other people doing the very high level stuff. And then we see that, oh, this for loop doesn't compute the right calculations. Can you do that? And then in another case, I might see that, OK, this other guy is really good at that. So I can be the general structure guy. So I can create these big files and try and kind of piece things together.

Others preferred simpler approaches, like pair-programming, or finding a way to split a workload into independent pieces:

Alfred: If you have smaller tasks that people can do individually, it's always nice to split up and then do each individual task instead. But if it's a more difficult thing to program and you really need to think about how you do the things you do, then pair programming is a really nice way of doing it.

Effective collaboration on larger projects depends on competence with the material practice of "Planning and Structuring"; however, this competence is not enough by itself. As discussed, some students found that the overarching structure of their code was best developed while coding it. In these cases, effective collaboration required students to work together iteratively, adapting to the changing needs of the code:

Andrew: I think it's much easier to collaborate on writing. Because then it's like, you take this part, I'll take this part. Whereas with code it's more like, they're tightly coupled. So that's one thing that's maybe missing here at physics. Having something about how to collaborate on code.

#### *4. Communicating results*

A central part of doing computational physics research is presenting research results using plots, figures, tables, and written explanations. This type of communication also requires attention to the audience; a close collaborator, for instance, will need little help to understand code and plots, while someone reading about one's work for the first time might need documentation, method descriptions, annotated and labeled results, and code examples.

For computational physics students at the journeyman level, plots are the primary mode for communicating results. Students begin developing these communication skills from the start of their education in physics, as plots are a deliverable for most of the computational physics exercises they solve:

Nils: I think one of the things that I learned the most from my bachelor's was how to try and visualize things and how to use like plots and different like Matplotlib to try and really tell the reader as much as possible in a very simple way.

Later in their studies, at the journeyman level, students learn additional plotting techniques, like scatter plots, heatmaps, and three-dimensional plots, as well as when to use them most effectively. As they gain experience with more open-ended problems, or problems with higher-dimensional or overwhelming amounts of data, students also learn to decide *what* to plot:

Nils: In the thesis, the amount of information that I tried to visualize just became bigger and bigger and these problems became more and more abstract. So I think one of the main things that I became hopefully even better at trying to figure out, okay, what is actually interesting to plot here? What are the metrics that I should try and visualize and how can I do this in a very pedagogical way?

Producing an effective plot or explanation requires some competence with the cognitive practice "Exploring and Interpreting," although more expertise is needed to use these plots as a communication tool, such as considering the needs and background of one's audience.

In some cases, the code itself was a shareable research result, and in these cases, students at the journeyman level would take into account additional considerations to help make the code easier to evaluate, expand, or reuse; for example, students might add explanations outside the code, like writing documentation, code examples, or computational notebooks:

Ivar: Especially when you deliver reports and stuff, you've often presented some results where you've used code to do it. And then it's very important that the code is well documented, so that people can come back through and reuse your code. At least that they can recreate the results that you've presented in some way. So what I often do in a way is to include my own code examples of how… that the code is well-documented, but in addition have a more explicit explanation of how the code has been used.

#### *5. Professionalized collaboration*

Although rare in our surveyed student sample, some students had experience collaborating in more structured, professional environments. These experiences often came through part-time work or through working in a formal research group. We place these elements at the top of the journeyman level of the social PCL pillar, as they represent a transition into higher-level expertise on collaboration and communication within computational physics research.

In these settings, students reported learning to adopt a more professional coding style in order to effectively work as part of the group. When contributing to these large long-term computational projects, learning to follow established conventions is important, as it makes it much easier to maintain their correctness, extensibility, and readability.

One student described their experience of learning to write code in the style of their research group's well-established framework. The student appreciated the quality of the work done in the group, and through guidance and practice, they learned to contribute to the project by imitating the established style:

Casey: When I write code in my supervisor's framework I make a conscious choice to do it the same way. I often take as a starting point how he has created methods and classes in that document here, and then I try to imitate. And it's partly because it should be good and partly that I don't want to introduce my bad habits, but also just because then there will be a continuity in the document that I think is more valuable than if I start doing something completely different that maybe could also be better, but it kind of becomes more messy then, so I try to imitate what he has done.

This approach to readability differs from the previously described readability practices, as it requires a level of consistency and humility not required when working on a typical assignment or independent master's project.

In professional research environments, the expectations of code quality are also higher, as code will likely be used by others over a longer time frame, and there is a greater need for robustness and reproducibility:

Nils: And for instance, now I know that my supervisors want me to, or I won't be able to write an article, but they want to kind of try and work further on many of these methods that I created. So they wanted me to kind of tidy everything.

Interestingly, although several students experienced these higher expectations from working with professionals, they also recognized the usefulness of quicker and messier coding when there was little need for code reuse or robust results.

## VII. DISCUSSION AND CONCLUSION

Returning to our first research question, "How do computational physics students develop their computational literacy to the journeyman level?," we have found that large computational projects play a critical role in this development. The size and complexity of such projects push students to learn how to structure, document, and modularize their code, and these projects offer authentic use-cases for object orientation, version control, and collaboration strategies. These skills mark a meaningful transition from the novice to the journeyman level of PCL.

We also found that part of the transition to the journeyman level comes from experience with professional computational physics research. This independent research, with its specialized tools and methods, along with its higher demands on robustness, drives students to learn specialized skills not gained from coursework. Additionally, we note the important role peers can play in cultivating PCL. Collaboration is important in computational science, and peers can give useful feedback about what works and what does not in a student's approach to collaborative coding. Peers can also recommend new tools or approaches to try out. Finally, students often relied on external resources like Stack Exchange, ChatGPT, and YouTube videos on programming to learn new techniques and resolve issues they encountered.

To answer our second research question, "what are the essential components of physics computational literacy at the journeyman level?," we have identified several essential elements of material, cognitive, and social PCL that go beyond what is learned in introductory physics and help students transition into computational physics research.

First, focusing on the material pillar of computational literacy, we see that knowledge of programming tools is both important and nontrivial at the journeyman level. At this level, the interviewed students were all fluent in at least one programming language and had little trouble with basic syntax or language features. They were also proficient with common coding tools such as code editors, external documentation, and AI systems. Efficient tool use is an integral part of physics computational literacy, as these tools make code faster to write and less prone to errors. Having access to a wider array of languages and packages, all with their own uses for computational physics, also increases the number of problems students can solve effectively.

In addition to being familiar with code and development tools, students at the journeyman level of PCL also have experience planning and structuring their programs in sophisticated ways, for instance using object orientation or creating modular code. Many students at this level described their coding approach as iterative. They preferred initial testing and exploration over planning out a detailed structure for complex problems. Students at this journeyman level are cognizant of when code needs to be fast and when it needs to be readable and documented. We found that some students even learned to use specialized tools for their thesis research, moving them closer to an expert level of PCL.

In comparing these results to the research literature, we see that this approach to programming matches well with the professional advice of Balaban *et al.* [34] to "start with prototypes and expand them in short development cycles" and "refactor frequently." Interviewed students also demonstrated other professional practices mentioned by Balaban *et al.* such as "avoid premature optimization" and "write self-documenting code for programmers." Notably, these practices were not found among the introductory students studied by Weller *et al.* [25], which adds evidence to our claim that they are characteristic of journeyman-level PCL. Other professional practices, like testing [35,36], were not as prevalent among the students in this study, likely because code and results produced for course assignments typically do not need to be especially robust or reproducible.

Second, focusing on the cognitive pillar, we found that students at the journeyman level of PCL had internalized a numerical view of physics, reflexively thinking of ways to discretize problems and solve them iteratively. This emphasis on a numerical view resonates with the importance of numerical analysis and representing data, also seen by Phillips *et al.* [32]. At this level, we also see the important role that visualization plays for the students' understanding and problem solving, also highlighted by Phillips's model. Building on these foundational skills, students at the journeyman level of PCL learn a large set of numerical methods, preparing them for the independent, highly specialized modeling done in professional research.

Third, focusing on the social pillar, we see that the students at the journeyman level of PCL learn to consider their code in relation to the code and practices of others. When collaborating, students at this level learn to structure, comment, document, and explain their code such that others can understand it. They are also able to delegate programming tasks and are cognizant of the challenges of dividing and recombining coding tasks. Some students learn more professional collaboration skills from working in research groups with established norms, where they encountered higher expectations for code reuse and documentation, two professional practices also highlighted by Balaban *et al.* [34].

As expected, we see significant interdependence and overlap between practices in the different pillars. For instance, the students' ability to delegate coding tasks relies on their ability to plan and structure code. In addition, communicating results relies on the ability to explore and interpret those results using visualization tools. Solving computational problems requires the requisite physics knowledge to develop an initial prototype into a full solution, as well as programming knowledge to know which aspects of the code will be easiest to implement. Optimization and programming tools also play an important part in the students' approach to modeling and in the specialized applications of computation in their thesis research. This interdependence makes clear that PCL beyond the introductory or novice level is best acquired holistically, through engagement with challenging tasks that require students to both leverage their existing PCL and further develop it [38].

We also see that the elements of the students' PCL are affected by the physics context. The elements in the cognitive pillar, of course, deal with the application of computation to physics problems directly. The students' programming toolkit and specialized tools are also dependent on the specific methods students are implementing. Additionally, the communication of results and professionalized collaboration that students learn are influenced by the norms and needs of the field. Though the rest of the PCL elements are not as clearly impacted by the physics context, we expect there to be nuanced differences in how physics students and other students learn and use these elements.

Based on these findings, we suggest several implications for teaching. First, as our data and previous research [9,32,38,52] suggest, computational projects are necessary to develop PCL to the journeyman level. Work on complex problems, closer to authentic research, allows students to cultivate more advanced skills, like structuring code, exploring complex data, and delegating interconnected coding tasks. The inherent challenge of large code, as well as the possibility to work on more complex physics problems, provides students with learning experiences that are not available when only working on short scripts or heavily scaffolded problems. Second, requiring students to work in groups on computational problems can be both useful and challenging. Group work can offer students valuable feedback on how to improve their coding and how to write code in a way that others will understand. However, as Burke and Atherton [9] found, group work can be challenging, and without guidance on how to divide labor and structure group work, students may be left with uneven shares of the project. Finally, instructors should familiarize themselves with the external resources available to assist students in learning programming and then make those resources easily available to all students. Here, it may be helpful to consult colleagues or computational physics researchers for suggestions on their go-to resources.

This study has several limitations that point the way to future work in this area. As mentioned, the student population in this study was limited, and all interviewed students came from the same institution and specialized degree program. Although this sample provides a uniquely rich perspective on the targeted phenomenon, our results may be less applicable to PCL for students who specialize in theory or experiments. Students in such contexts, for example, might develop more expertise with the interface between computation and scientific equipment or how to use computational tools to build theoretical and mathematical understanding of problems. Professional laboratories often include a combination of computational and "non-computational" experts, which presents different challenges than groups in which all participants are proficient coders, as was the case for the study participants. Additionally, the students in this study were a group who had a clear interest in computational physics, even choosing the subject for their master's degree; this limitation is important in considering our findings on the role of courses and less-formal learning environments, as students with lower levels of interest or background in coding might experience these learning environments differently. Finally, we note that the retrospective nature of these interviews means that we are relying on students' self-reported attitudes and perspectives (albeit triangulated with examples of their research work), which means we lack some detail and certainty about how these students work on computational problems in the moment.

We therefore see a need to study the development and composition of PCL in other institutions and programs, with different goals and approaches to teaching computation, in order to see the effects of this variation on students' development of physics computational literacy. These types of studies would give us valuable insight into strategies for educating students in computational physics as compared to theoretical or experimental physics, or how students develop from journeymen to experts. We also see a need to study students' in-the-moment problem solving and collaboration in computational physics environments beyond the introductory level to add more specificity and detail to these results. The impact of generative AI on computational literacy will also require further research, as most of the students in this study only had limited access to such tools at the time of data collection and thus had not deeply integrated them into their workflows. In the present day, we hypothesize that students likely integrate AI to a much greater degree into both the way they learn and do computational physics; however, research is needed on the different ways students work productively with AI and what effects this has on their development of physics computational literacy.

In conclusion, computation has become a crucial pillar of physics, standing alongside theory and experiment as the cornerstones of our field. As this technology, and its use in physics, continues to develop, it is vital to study and improve the ways we help students become computationally literate in physics to best prepare the next generation of researchers.

## ACKNOWLEDGMENTS

This work was funded by the Norwegian Directorate for Higher Education and Skills (HK-dir), which supports the University of Oslo's Center for Computing in Science Education.

## DATA AVAILABILITY

The data that support the findings of this article are not publicly available because they contain sensitive personal information. The data are available from the authors upon reasonable request.